\documentclass[11pt]{article}
\usepackage{latexsym}
\usepackage{fullpage}
\usepackage{boxedminipage}

\newtheorem{theorem}{Theorem}

\usepackage{rotating}
\usepackage{url}
\usepackage{fancyhdr}

\fancypagestyle{plain}{}

\begin{document}

\setcounter{page}{81}

\title{Failure of A Mix Network}

\author{Kun Peng\\
Institute for Infocomm Research\\
dr.kun.peng@gmail.com}
\date{}

\maketitle

\begin{abstract}

A mix network by Wikstrom fails in correctness, provable privacy and soundness. Its claimed advantages in security and efficiency are compromised. The analysis in this paper illustrates that although the first two failures may be fixed by modifying the shuffling protocol, the last one is too serious to fix at a tolerable cost. Especially, an attack is proposed to show how easily soundness of the shuffling scheme can be compromised. Moreover, the most surprising discovery in this paper is that it is formally illustrated that in practice it is impossible to fix soundness of the shuffling scheme by Wikstrom. 

\end{abstract}

\section{Introduction}
\label{intr}

Shuffling is a very important cryptographic technique to build mix network, which is popular kind of anonymous communication channel. The most important application of mix network is electronic voting, which is a highly sensitive application with critical requirements on security. As large-scale election applications may involve a very large number of voters, high efficiency is often desired in e-voting.
In recent years, a few shuffling-based mix network schemes \cite{Abe1999,Abe2001,Furukawa2001,Neff2001,Groth2003,Neff2004,Peng2004A,Furukawa2005,Peng2005A,Peng2011A,Nguyen2006,Wikstrom2005,Groth2007,Groth2008} have been proposed. They claim to achieve strong security and high efficiency.
Among them, the most efficient in computation\footnote{Like most work in shuffling, we focus on computation when discussing efficiency as in most cases communicational cost is a monotone function of computational cost. The only exception is \cite{Groth2008}, which sacrifices computational efficiency to achieve high efficiency in communication.} are \cite{Peng2004A}, \cite{Groth2007} and \cite{Wikstrom2005}.
However, all of these three schemes have weakness in security, although with difference in kind and degree. 

As explained in Appendix~\ref{dr}, the shuffling schemes in \cite{Peng2004A} and \cite{Groth2007} have some drawbacks, which make them unsuitable for applications with very critical security requirements like political e-voting.
Although the drawbacks of \cite{Peng2004A} and \cite{Groth2007} limit their application, 
if they are not applied to applications with very high security requirements, they are still very useful in practice.  
As a result, if it is as secure as it claims, the shuffling scheme in \cite{Wikstrom2005} is the most efficient solution in computation still remaining secure for shuffling based e-voting. 
An important question is: is the shuffling scheme in \cite{Wikstrom2005} as secure as it claims?
Our analysis demonstrates that the security problems in \cite{Wikstrom2005} are much more serious than in \cite{Peng2004A} and \cite{Groth2007} and compromise the most important security properties. 

Our analysis start with a more obvious but less serious problem: 
the shuffling scheme in \cite{Wikstrom2005} fails in correctness (defined in Section~\ref{bac}) and its proof of zero knowledge is based on incorrect operations. As a result, to fit the proof of zero knowledge, a valid shuffling operation must risk a failure in verification.
Our analysis illustrates that due to two reasons, keeping the shuffling scheme unchanged and decreasing the probability of failure of correctness by using special parameter setting is \textbf{not} an effective method to improve security.
Firstly, there is a dilemma between correctness and provable zero knowledge and decreasing the probability of failure of correctness will increase the probability of failure of proof of zero knowledge. 
Secondly and more importantly, as explained later the same inappropriate operations compromising correctness prevent the author's suggested method from implementing soundness as well.
An appropriate modification can achieve correctness and at the same time make implementation of soundness as suggested by the author possible (although not necessarily guaranteeing soundness).
So a more comprehensive countermeasure is needed and thus a modification is proposed in this paper to fix the shuffling scheme in \cite{Wikstrom2005} and achieve correctness. As this modification affects provability of zero knowledge, a new privacy analysis is needed to demonstrate achievement of statistical zero knowledge in the modified shuffling protocol.
As explained later, this modification is necessary in trying to implement author's suggested method to achieve soundness.

\fancyhead{}
\fancyfoot{}

\fancyhead[CO,CE]{International Journal of Network Security \& Its Applications (IJNSA), Vol.3, No.1, January 2011}
\fancyfoot[RO,RE]{\thepage}

A more serious, more fatal but less obvious problem of \cite{Wikstrom2005} lies in soundness (defined in Section~\ref{bac}). 
A key technique to guarantee soundness of the shuffling scheme in \cite{Wikstrom2005} is \textit{range proof}, which must be run multiple times to guarantee that multiple integers are in a special range. However, range proof is not implemented in \cite{Wikstrom2005}. On one hand, in the outline of the shuffling scheme in \cite{Wikstrom2005}, range proof is emphasize to be necessary; on the other hand, in the detailed and complete implementation of the shuffling scheme, range proof is not implemented or even mentioned in any of the multiple operations depending on it for soundness. 
So there is a mystery: is range proof in \cite{Wikstrom2005} too simple and too straightforward so ignored in the implementation or too difficult and too controversy so avoided in the implementation? 
As it determines soundness of the shuffling scheme in \cite{Wikstrom2005}, the most efficient still-surviving shuffling scheme, this question must be answered. 
However, as the information about the details of range proof in \cite{Wikstrom2005} is only a vague sentence, it is not easy to figure the question out.
Too sketchy description of key operation leaves a few doubts for the readers to clear.
Firstly, is the author really aware the necessity of range proof? How important is range proof? Can it be ignored?
Secondly, can the range proof primitive suggested by the author implement range proof in his shuffling scheme? Can we ask the author to give an implementation with detailed parameter setting and operations? If his method fails, does it just cannot work or need optimisation in details (e.g. parameter setting and operational details)? Can we adjust the parameter setting or operational details to fix the problem? If the adjustment is difficult, can we formally prove that his method is irremediable and cannot be fixed at all?
Thirdly, even if the author's range proof cannot work at all, can it be replaced by other range primitives unknown to the author? Can another technique implement range proof in the shuffling scheme in \cite{Wikstrom2005} at a tolerable cost?
To discover the truth, we need to answer all the questions.

When the given information is not enough and many possibilities are left, the most reliable way to handle it is to explore and try every possibility. If every possible detailed implementation for a method is shown to not work, the method must fail. When there are a few doubts about a scheme, the most responsible method to assess it is to discuss and answer the doubts one by one. More doubts are verified to be problems, more sure are the readers about failure of the scheme.
In this way, we can attack a scheme and demonstrate its failure in the most convincing way, leaving no space for any excuse or defense, even if its problems are hidden in incomplete, vague and sketchy descriptions.
Firstly, an attack is proposed to completely compromise soundness of the shuffling scheme in \cite{Wikstrom2005} in absence of range proof. The attack allows invalid shuffling to pass the verification of the shuffling protocol in \cite{Wikstrom2005}, so the shuffled messages can be tampered with without being detected.
So its missing in the detailed implementation is an inexcusable mistake. If the implementation cannot be provided, the shuffling scheme definitely fail in soundness. 
Secondly, it is clearly demonstrated that the range proof primitive suggested by the author cannot implement range proof either in his original shuffling scheme or in the shuffling scheme modified by us to achieve correctness and make implementation of soundness possible\footnote{As discussed before, the original shuffling scheme in \cite{Wikstrom2005} is inconsistent with the author's suggested method to implement soundness, while the shuffling scheme modified by us to achieve correctness avoids the contradiction although not necessarily guaranteeing soundness.}.
Further analysis formally illustrates that the range proof primitive suggested in \cite{Wikstrom2005} is irremediable and always violates soundness of the shuffling scheme no matter how it adjusts parameters and details. So the problem in soundness of \cite{Wikstrom2005} is not only incompleteness in the claimed complete and detailed implementation, but also an irremediable mistake of unimplementability of key operation, which cannot be fixed at all and so cannot be excused as carelessness in implementation details.
Thirdly, every other range proof primitive not mentioned by the author is explored and each of them is demonstrated to greatly deteriorate efficiency of the shuffling scheme in \cite{Wikstrom2005} and turn it into one of the least efficient shuffling schemes if being employed.
Thus, the final chance to maintain security and efficiency of the shuffling scheme in \cite{Wikstrom2005} is eliminated.
Therefore, it is concluded beyond any doubt and excuse that soundness not only fails but also cannot be fixed at a tolerable cost in the shuffling scheme in \cite{Wikstrom2005}, which is the \textbf{most important contribution and most surprising discovery} in this paper. 

\section{Background: the Shuffling Scheme in \cite{Wikstrom2005}}
\label{bac}

In a shuffling protocol, a shuffling node re-encrypts and reorders multiple input ciphertexts to some output ciphertexts such that the messages encrypted in the output ciphertexts are a permutation of the messages encrypted in the input ciphertexts.
Shuffling is usually employed to build up anonymous communication channels and its most important application is e-voting.
The following properties must be satisfied in a shuffling protocol.
\begin{itemize}
	\item Correctness: if the shuffling node strictly follows the shuffling protocol, the shuffling protocol ends successfully and the plaintexts encrypted in the output ciphertexts are a permutation of the plaintexts encrypted in the input ciphertexts.
	\item Public verifiability: the shuffling node can publicly prove that he does not deviate from the shuffling protocol.
	\item Soundness: a successfully verified proof by a shuffling node guarantees that the plaintexts encrypted in the output ciphertexts are a permutation of the plaintexts encrypted in the input ciphertexts without any trust assumption on the shuffling node.
	\item Zero knowledge (ZK) Privacy: The permutation used by the shuffling node is not revealed. More formally, a simulating transcript indistinguishable from the real shuffling transcript can be generated by a polynomial party without any knowledge of the shuffling node's secret inputs.
\end{itemize} 

Shuffling is frequently employed in anonymous communication and its most important application is electronic voting, where the voters need to anonymously cast their votes. 
As stated in Section~\ref{intr}, the shuffling scheme in \cite{Wikstrom2005} is a very efficient solution to shuffling based e-voting. Its main idea is simple. Suppose $N$ ElGamal ciphertexts $(u_1,v_1)$, $(u_2,v_2)$, $\dots, ~(u_N,v_N)$ are input to a shuffling node, which then outputs ciphertexts $(u'_1,v'_1)$, $(u'_2,v'_2)$, $\dots, ~(u'_N,v'_N)$. To prove that the messages encrypted in $(u'_1,v'_1)$, $(u'_2,v'_2)$, $\dots, ~(u'_N,v'_N)$ is a permutation of the messages encrypted in $(u_1,v_1)$, $(u_2,v_2)$, $\dots, ~(u_N,v_N)$, given random primes $p_1,p_2,\dots,p_N$ in $[2^{K_3-1},2^{K_3}-1]$ ($[~]$ stands for a range of consecutive integers as defined in \cite{Wikstrom2005}), the shuffling node only needs to prove that he knows secret integers $\rho_1,\rho_2,\dots,\rho_N$ and $\pi()$, a secret permutation of $1,2,\dots,N$, to satisfy
\begin{eqnarray}
\label{e1}
& \prod_{i=1}^N(D(u_i,v_i))^{p_i}=\prod_{i=1}^N(D(u'_i,v'_i))^{\rho_i}, \\
\label{e2}
& \rho_i=p_{\pi(i)} \mbox{ for } i=1,2,\dots,N.
\end{eqnarray}
where $D()$ stands for decryption.
For privacy of the shuffling, neither any $\rho_i$ nor $\pi()$ can be revealed in the proof.
It has been illustrated in \cite{Groth2003} that (\ref{e1}) and (\ref{e2}) guarantee that the messages encrypted in $(u'_1,v'_1)$, $(u'_2,v'_2)$, $\dots, ~(u'_N,v'_N)$ is a permutation of the messages encrypted in $(u_1,v_1)$, $(u_2,v_2)$, $\dots, ~(u_N,v_N)$. Soundness of this idea is more formally proved in \cite{Peng2005A,Peng2011A}. For all the shuffling schemes employing this idea, satisfaction of (\ref{e1}) is easy to prove and the key technique is how to prove satisfaction of (\ref{e2}). The method to prove (\ref{e2}) is claimed to be more efficient in \cite{Wikstrom2005} than in \cite{Groth2003} and \cite{Peng2005A,Peng2011A}. Satisfaction of (\ref{e2}) is reduced to satisfaction of the following three equations in \cite{Wikstrom2005} where choice of $K$ and its relation to other parameters are absent in \cite{Wikstrom2005} and will be discussed later in Section~\ref{imp}.
\begin{eqnarray}
\label{e8.5}
& -2^K+1\le \rho_i\le 2^K-1 \mbox{ for } i=1,2,\dots,N\\
\label{e9}
& \prod_{i=1}^Np_i=\prod_{i=1}^N\rho_i \\
\label{e10}
& \sum_{i=1}^Np_i=\sum_{i=1}^N\rho_i  
\end{eqnarray}

A detailed proof protocol called Protocol 2 is employed in \cite{Wikstrom2005} to implement proof of (\ref{e8.5}), (\ref{e9}) and (\ref{e10}). 
Note that Protocol 2 is not a sketchy outline but supposed to be a complete implementation with every detail of the shuffling scheme in \cite{Wikstrom2005}.
Proof of (\ref{e9}) and (\ref{e10}) is a straightforward application of zero knowledge proof of equality of discrete logarithms \cite{Chaum1992}, so quite easy. The key technique is proof of satisfaction of (\ref{e8.5}). As Protocol 2 in \cite{Wikstrom2005} is a quite complex 7-step proof protocol, it is not recalled here in its fully complete form and interested readers can find its fully complete description in \cite{Wikstrom2005}. However, the operations closely related to proof and verification of (\ref{e8.5}) in Protocol 2 in \cite{Wikstrom2005} is extracted as follows where definition of all the involved integers can be found in \cite{Wikstrom2005}.
\begin{itemize}
	\item In Step 6 of Protocol 2 in \cite{Wikstrom2005}, the prover calculates and publishes 
\begin{eqnarray}
\label{o1}
& e_i=ct_i+s_i \bmod 2^{K_2+K_4+2K_5} \\
\label{o2}
& e'_i=ct'_i+s'_i \bmod 2^{K_2+K_4+2K_5} \\
\label{o3}
& d_i=cp_{\pi(i)}+r_i\bmod 2^{K_3+K_4+K_5} \\
\label{o4}
& e=ct+s \bmod 2^{K_2+NK_3+K_4+K_5+\log_{2}N} \\
\label{o5}
& e'=ct'+s' \bmod 2^{K_2+K_5+\log _{2}N}  
\end{eqnarray}
where $K_2,K_3,K_4,K_5$ are integers defined in \cite{Wikstrom2005} as security parameters.
	\item In Step 7 of Protocol 2 in \cite{Wikstrom2005}, it is verified
\begin{eqnarray}
\label{e'0}
& \textbf{b}_i^c\gamma_i=\textbf{h}^{e_i}\textbf{b}_{i-1}^{d_i} \\
\label{e'0.5}
& {\textbf{b}'}_i^c\gamma'_i)=\textbf{h}^{e'_i}\textbf{g}^{d_i} \\
\label{e'1}
& (b_1^c\alpha_1,~(V/b_2)^c\alpha_2,~W^c\alpha_3)= \\
& (g^{f_1}\prod_{i=1}^N(u'_i)^{d_i},~g^{-f_2}\prod_{i=1}^N(v'_i)^{d_i},~g^{f_{r'}}\prod_{i=1}^N(g_i)^{d_i}) \nonumber\\
\label{e'2}
& (\textbf{b}_i^c\gamma_i,~({\textbf{b}'}_i^c\gamma'_i)=(\textbf{h}^{e_i}\textbf{b}_{i-1}^{d_i},~\textbf{h}^{e'_i}\textbf{g}^{d_i}) \\
\label{e'3}
& (\textbf{g}^{-\prod_{i=1}^Np_i}\textbf{b}_N)^c\gamma=\textbf{h}^e \\
\label{e'4}
& (\textbf{g}^{-\sum_{i=1}^Np_i}\prod_{i=1}^N\textbf{b}'_i)^c\gamma=\textbf{h}^{e'}
\end{eqnarray} 
\end{itemize}

It is claimed in \cite{Wikstrom2005} that the shuffling protocol as described in Protocol 2 is a complete implementation to achieve correctness, soundness and zero knowledge in privacy.

\section{Correctness and Provable Zero Knowledge --- Failure and Fixing}
\label{ff}

In this section correctness and provability of zero knowledge of the shuffling protocol in \cite{Wikstrom2005} are shown to fail as key operations employ wrong moduli. 
Our analysis illustrates that due to two reasons, keeping the wrong moduli and decreasing the probability of failure of correctness by using special parameter setting is not an effective method to improve security.
Firstly, there is a dilemma between correctness and provable zero knowledge and decreasing the probability of failure of correctness will increase the probability of failure of proof of zero knowledge. 
Secondly and more importantly, as explained later the wrong moduli prevent the author's suggested method from implementing soundness as well.
So a modification is proposed to achieve correctness, while provability of zero knowledge and soundness are taken into account. We have to emphasize that it is still needed to prove achievement of statistical zero knowledge in a new proof method. As such needed proof or argument in statistical sense have been used in similar circumstances \cite{Poupard2000,Boudot1999,Shoup1999,Camenisch2001} and can be adopted in the shuffling protocol in \cite{Wikstrom2005}, they are not detailed in this paper due to space limitation. 
As explained later, this modification is necessary in trying to implement author's suggested method to achieve soundness.

\subsection{Failure of Correctness}

Correctness of the shuffling scheme in \cite{Wikstrom2005} requires that if the shuffling node strictly follows the shuffling protocol and does not deviate from it in any way, he can pass all the verifications in Step 7 of Protocol 2 in \cite{Wikstrom2005}. 
However, satisfaction of (\ref{e'0}), (\ref{e'0.5}), (\ref{e'1}), (\ref{e'2}), (\ref{e'3}) and (\ref{e'4}) is not guaranteed in \cite{Wikstrom2005} even if the shuffling node strictly follows the shuffling protocol and does not deviate from it.
More precisely, although (\ref{e'0}), (\ref{e'0.5}), (\ref{e'1}), (\ref{e'2}), (\ref{e'3}) and (\ref{e'4}) are satisfied when 
\begin{eqnarray}
\label{n1}
& e_i=ct_i+s_i \mbox{ in } Z,  \\
\label{n2}
& e'_i=ct'_i+s'_i \mbox{ in } Z,  \\
\label{n3}
& d_i=cp_{\pi(i)}+r_i \mbox{ in } Z,  \\
\label{n4}
& e=ct+s \mbox{ in } Z,  \\
\label{n5}
& e'=ct'+s' \mbox{ in } Z,  
\end{eqnarray} 
their satisfaction are not guaranteed when (\ref{o1}), (\ref{o2}), (\ref{o3}), (\ref{o4}) and (\ref{o5}) are employed in Step 6 of Protocol 2 in \cite{Wikstrom2005} as
\begin{itemize}
	\item the order of $\textbf{h}$ is $(\textbf{p}-1)(\textbf{q}-1)/2$ instead of $2^{K_2+K_4+2K_5}$ and $ct_i+s_i$, $ct'_i+s'_i$ distribute beyond $2^{K_2+K_4+2K_5}$;
	\item the order of $g_i$ is $q$ as set in Section 2 of \cite{Wikstrom2005} instead of $2^{K_3+K_4+K_5}$ and $cp_{\pi(i)}+r_i$ distributes beyond $2^{K_3+K_4+K_5}$;
	\item the parameter setting of the encryption algorithm in Section 4 of \cite{Wikstrom2005} implies that the order of $u'_i$ is $q$ instead of $2^{K_3+K_4+K_5}$ and $cp_{\pi(i)}+r_i$ distributes beyond $2^{K_3+K_4+K_5}$; 
	\item in Section 4.5 of \cite{Wikstrom2005}, the author assumes $m_i\in G_q$, so the order of $v'_i$ is $q$ instead of $2^{K_3+K_4+K_5}$ and $cp_{\pi(i)}+r_i$ distributes beyond $2^{K_3+K_4+K_5}$; 
	\item the order of $\textbf{h}$ is secret and not $2^{K_2+NK_3+K_4+K_5+\log_{2}N}$ and $ct+s$ distributes beyond $2^{K_2+NK_3+K_4+K_5+\log_{2}N}$;
	\item the order of $\textbf{h}$ is secret and not $2^{K_2+K_5+\log_{2}N}$ and $ct'+s'$ distributes beyond $2^{K_2+K_5+\log _{2}N}$.
\end{itemize}
where $\textbf{p}$, $\textbf{q}$ and $q$ are secret system parameters defined in \cite{Wikstrom2005} and cannot be used by the prover.
In Section~\ref{pzk}, Section~\ref{fc} and Section~\ref{imp}, it is illustrated that the problem in correctness cannot be solved by reducing the probability that correctness fails.

\subsection{Dilemma between Correctness and Proof of Zero Knowledge}
\label{pzk}

Failure of correctness seems to be a careless mistake and easy to fix. Removing the moduli when calculating $e_i,~e'_i,~d_i,~e,$ and $e'$ will lead to complete correctness. 
Alternatively, setting the parameters with appropriate values can make the probability that the moduli are used in calculating the five integers negligible and thus guarantee correctness with a large probability.
However, it is not so simple. Let's see why moduli different from the orders of the responses are employed. When the orders are unknown, isn't the non-modulus calculation simpler and completely consistent with correctness?
The reason is that the moduli are employed and they need take effect in calculation of the responses with a large probability as strict and formal zero knowledge is desired in \cite{Wikstrom2005}.
So both these two countermeasures contradict proof of zero knowledge in \cite{Wikstrom2005}.
Zero knowledge of Protocol 2 in \cite{Wikstrom2005} is proved in Proposition 1 in D.1 in Page 31 of \cite{Wikstrom2005B} (2005 Version), in which an explicitly emphasized necessary condition for zero knowledge of Protocol 2 is that 
\begin{itemize}
	\item $e_i=ct_i+s_i \bmod 2^{K_2+K_4+2K_5}$ such that $e_i$ is uniformly distributed in $Z_{2^{K_2+K_4+2K_5}}$ and thus can be simulated;
	\item $e'_i=ct'_i+s'_i \bmod 2^{K_2+K_4+2K_5}$ such that $e'_i$ is uniformly distributed in $Z_{2^{K_2+K_4+2K_5}}$ and thus can be simulated;
	\item $d_i=cp_{\pi(i)}+r_i\bmod 2^{K_3+K_4+K_5}$ such that $d_i$ is uniformly distributed in $Z_{2^{K_3+K_4+K_5}}$ and thus can be simulated;
	\item $e=ct+s \bmod 2^{K_2+NK_3+K_4+K_5+\log_{2}N}$ such that $e$ is uniformly distributed in $Z_{2^{K_2+NK_3+K_4+K_5+\log_{2}N}}$ and thus can be simulated;
	\item $e'=ct'+s' \bmod 2^{K_2+K_5+\log _{2}N}$ such that $e'$ is uniformly distributed in $Z_{2^{K_2+K_5+\log _{2}N}}$ and thus can be simulated.
\end{itemize}
So the modulo computations in (\ref{o1}), (\ref{o2}), (\ref{o3}), (\ref{o4}) and (\ref{o5}) are deliberately used for the sake of proof of zero knowledge in \cite{Wikstrom2005}. 
Both the two countermeasures calculate the five integers without any modulus with at least an overwhelmingly large probability. In this case, their distribution is not uniform as claimed and needed in Proposition 1 in \cite{Wikstrom2005}. Instead, their distribution is more dense in the middle of their distribution range and more sparse near the edge of their distribution range.
So the proof of Proposition 1 in \cite{Wikstrom2005} fails and new proof of zero knowledge is needed.
Actually, with both the two countermeasures the five integers become monotone functions of five corresponding secret integers with at least an overwhelmingly large probability, so publication of them reveals some information about the secret integers and thus (at least partially) compromises the claimed zero knowledge property.
This dilemma will be solved in Section~\ref{fc}.

\subsection{Fixing the Two Drawbacks}
\label{fc}

To fixing the two drawbacks, we only have the following two options, either keeping the wrong moduli or removing them. 
\begin{itemize}
	\item Option 1:\\
	Still employing (\ref{o1}), (\ref{o2}), (\ref{o3}), (\ref{o4}) and (\ref{o5}) with the wrong moduli
in Step 6 of Protocol 2 in \cite{Wikstrom2005} and relying on its Proposition 1 for zero knowledge, with a hope that the probability of failure of correctness is low or even negligible without compromising proof of zero knowledge. This hope is unrealistic as 
\begin{itemize}
\item if the modulus computation takes effect with a large probability when a response is calculated as it otherwise overflows the modulus, correctness of the shuffling scheme fails;	
\item if the modulus computation does not take effect with a large probability when a response is calculated as it is small enough, proof of Proposition 1 and thus zero knowledge property in \cite{Wikstrom2005} fail as explained in Section~\ref{pzk}.
\end{itemize}
So with this option, correctness and provable zero knowledge cannot be achieved simultaneously and at least one of them must fail in \cite{Wikstrom2005} no matter how parameter setting is adjusted. 
	\item Option 2\\
Since modulo operations are expected to take effect with a negligible probability for the sake of security, why not remove them? Employing the modified operations
		(\ref{n1}), (\ref{n2}), (\ref{n3}), (\ref{n4}) and (\ref{n5})
in Step 6 of Protocol 2 in \cite{Wikstrom2005}, abandoning its Proposition 1 and designing a new proof mechanism to demonstrate zero knowledge. In doing this, the following difficulties must be noticed.
\begin{itemize}
	\item As the responses are calculated without any modulus and become monotone functions of the corresponding secrets, when the secrets are unknown they cannot be simulated without any difference. So proof of ZK must be upgraded.
	\item The new zero knowledge proof is more complex as it involves statistical ZK \cite{Poupard2000,Boudot1999,Shoup1999,Camenisch2001}, which proves two distributions are different but cannot be distinguished. 
	That may be the reason why the modulo computations are still employed in calculation of the responses in \cite{Wikstrom2005} although they compromise correctness.
\end{itemize}
\end{itemize}
 
A comprehensive solution is designed based on Option 2. Namely, the modified operations (\ref{n1}), (\ref{n2}), (\ref{n3}), (\ref{n4}) and (\ref{n5}) are adopted in Step 6 of Protocol 2 in \cite{Wikstrom2005} and the modified protocol is called MP2 (modified protocol 2). This choice is made due to two reasons.
Firstly, Option 1 cannot handle the dilemma between correctness and provable zero knowledge in \cite{Wikstrom2005}.   
Secondly, as discussed in Section~\ref{imp}, Option 1 is inconsistent with the author's suggested method to implement soundness.
As mentioned before, MP2 only achieves statistical zero knowledge and thus modeling and analysis of ZK privacy of the shuffling scheme must be completely upgraded using statistical zero knowledge techniques.
Fortunately, proof of achievement of statistical zero knowledge in the case of MP2 is a mature technique and has been specified and explained very clearly in the literature \cite{Poupard2000,Boudot1999,Shoup1999,Camenisch2001}. So due to space limit, its details are not provided here and interested readers can read the literature.

\section{A More Serious Problem: Failure and Infeasibility of Soundness}

In this section, soundness of the shuffling scheme in \cite{Wikstrom2005} is demonstrated to fail. A very important operation is missing and cannot be implemented as suggested. The method suggested for the implementation is formally illustrated to fail and to be irremediable. The efficiency claim in \cite{Wikstrom2005} implies that no alternative method (e.g. \cite{Boudot2000} or \cite{Lipmaa2003A}) is efficient enough for the implementation. The most important and surprising discovery is that we can formally prove that no modification or optimisation can satisfactorily fix the problem in soundness.

To achieve soundness in the shuffling scheme in \cite{Wikstrom2005}, Protocol 2 as a complete implementation of it must guarantee that (\ref{e8.5}), (\ref{e9}) and (\ref{e10}) are satisfied. Satisfaction of (\ref{e9}) and (\ref{e10}) is straightforward, but we find no way to satisfy (\ref{e8.5}) in Protocol 2 in \cite{Wikstrom2005}. Actually, parameter $K$ is not mentioned in any way in Protocol 2 in \cite{Wikstrom2005}, which is $K$ free and thus (\ref{e8.5}) free. Then how can Protocol 2 in \cite{Wikstrom2005} guarantee (\ref{e8.5})? We test Protocol 2 in \cite{Wikstrom2005} and find that as long as (\ref{e9}) and (\ref{e10}) are satisfied the prover can always pass the verification of Protocol 2 in \cite{Wikstrom2005} no matter whether $-2^K+1\le \rho_i\le 2^K-1$ is satisfied for any chosen $K$. Actually, once (\ref{e9}) and (\ref{e10}) are satisfied, no additional requirement on any $\rho_i$ (e.g. requiring it to be in $[-2^K+1,2^K-1]$ or any other range) is needed to pass the verification in Protocol 2 in \cite{Wikstrom2005}. 
To more convincingly demonstrate vulnerability of Protocol 2, an attack is proposed to compromise soundness of  in \cite{Wikstrom2005}.
Note that Protocol 2 is not a sketchy outline but supposed to be a complete implementation with every detail of the shuffling scheme in \cite{Wikstrom2005}. So there is a problem of incompleteness in implementation.

Our analysis demonstrates that the problem in range proof in \cite{Wikstrom2005} is not only incompleteness in the claimed detailed and complete implementation, but also a complete failure of range proof suggested by the author.  
To confirm the author's suggestion, a comprehensive analysis of available tools is employed to clarify a vague and sketchy hint, which may have hidden the problem. Our method is to have a comprehensive survey of every possibility and eliminate each infeasible choice, so that the only possible choice can be found, which mostly fits the author's suggestion and meets the efficiency claim but cannot work.
This method finds that although the original shuffling scheme in \cite{Wikstrom2005} is inconsistent with the author's suggested method to implement soundness, there is no contradiction between MP2 and the author's suggested method to implement soundness\footnote{This discovery has been mentioned in Section~\ref{intr}, Section~\ref{ff} and Section~\ref{fc} and will be detailed in Section~\ref{imp} as a reason for modifying the original shuffling scheme in \cite{Wikstrom2005} into MP2.}.
Unfortunately, it is then formally illustrated that even if range proof is implemented as suggested in \cite{Wikstrom2005} in MP2 it cannot guarantee satisfaction of (\ref{e8.5}).
Moreover, it is formally illustrated that the problem cannot be fixed by modifying the range proof technique (e.g. adjusting parameters or other details). In addition, alternative range proof techniques (which are not considered or mentioned in \cite{Wikstrom2005}) cannot fit the requirements on security and efficiency of \cite{Wikstrom2005}. Therefore, beyond any doubt and excuse soundness not only fails but also cannot be fixed at a tolerable cost in the shuffling scheme in \cite{Wikstrom2005}.

\subsection{An Attack against Soundness}

When (\ref{e8.5}) is not satisfied, satisfaction of (\ref{e2}) cannot be guaranteed by only (\ref{e9}) and (\ref{e10}). Therefore, in the shuffling scheme in \cite{Wikstrom2005}, as (\ref{e8.5}) is not guaranteed the messages encrypted in the output ciphertexts are not guaranteed to be a permutation of the messages encrypted in the input ciphertexts and thus its soundness fails.
More precisely, when the sum of $\rho_i$s equals the sum of $p_i$s and the product of $\rho_i$s equals the product of $p_i$s,
there is no guarantee that $\rho_i$s is a permutation of $p_i$s.
When only (\ref{e9}) and (\ref{e10}) are satisfied, a simple attack and be launched so that $\rho_i$s is not a permutation of $p_i$s.
A shuffling node chooses some $\rho_i$ as the products of multiple $p_i$'s and 1 or -1 and some other $\rho_i$ as 1 or -1 such that the sum of $\rho_i$s equals the sum of $p_i$s, just so simple.
There are many such choices for the attack to succeed.
A simple example is $N=10$, $p_1=p_2=\dots=p_{10}=2$ while $\rho_1=\rho_2=\rho_3=\rho_4=4$, $\rho_5=\rho_6=2$, $\rho_7=\rho_8=1$ and $\rho_9=\rho_{10}=-1$.
Another simple examples is $N=10$, $p_1=p_2=2$, $p_3=p_4=3$, $p_5=p_6=5$, $p_7=p_8=7$, $p_9=p_10=11$ while $\rho_1=\rho_2=22$, $\rho_3=\rho_4=15$, $\rho_5=\rho_6=-7$, $\rho_7=\rho_8=\rho_9=\rho_{10}=-1$.

Readers can easily verify that these two examples can pass the verification operations in \cite{Wikstrom2005} while $\rho_i$s is not a permutation of $p_i$s.
We checked each of the three membership tests and the eight equations in the verification (Step 7) of Protocol 2 in \cite{Wikstrom2005} and found that none of them can detect this attack. So this attack can pass the verification of the shuffling scheme in \cite{Wikstrom2005}.
Although the on-line version of \cite{Wikstrom2005B} is modified recently to defend its mistake, concrete counter-examples are much more convincing than its argument.
We wrote a computer program to search for such examples given $p_1,p_2,\dots,p_N$ and found countless of them, each of which is a counterexample to Protocol 2 in \cite{Wikstrom2005}.
Although our program uses brutal-force search, the search is still fast on a normal desktop when $N$ is not large. 
When the attack works with an $N_1$, it works as well with a larger $N_2$ as putting an incorrect permutation of $N_1$ ciphertexts and a correct permutation of $N_2-N_1$ ciphertexts together produces an incorrect permutation of $N_2$ ciphertexts.
So the attack can always be efficient no matter how large $N$ is. 
It has been illustrated in \cite{Groth2003,Peng2005A,Peng2011A} that when the $\rho_i$s is not a permutation of the $p_i$s, incorrect shuffling can satisfy (\ref{e1}) and pass the verification and thus soundness of shuffling is broken. So this attack compromises soundness of shuffling in \cite{Wikstrom2005}.

This attack convincingly shows how serious the crisis of soundness is when proof of (\ref{e8.5}) is not implemented.
So there is a question: in \cite{Wikstrom2005}, as a very important operation, is proof of (\ref{e8.5}) omitted due to the author's carelessness or just infeasible to implement as efficiently as claimed?

\subsection{Comprehensive Analysis to Clarify the Author's Suggestion} 
\label{imp}

Last subsection convincingly demonstrates the necessity of (\ref{e8.5}) with a concrete countermeasure.
Note that (\ref{e8.5}) contains $N$ instances of range proof, in which a prover commits to (or encrypt) an integer and then proves that the committed integer is in a certain range consisting of some consecutive integers.
If the author had thought range proof is too simple and straightforward in his shuffling scheme and thus had ignored it or he had been careless and forgotten to implement it, the range proof technique suggested by him would be able to implement range proof in his shuffling scheme. 
The only clue about range proof in \cite{Wikstrom2005} is ``We then note that a standard proof of knowledge over a group of unknown order also gives an upper bound on the bit-size of the exponents, i.e., it implicitly proves that $\rho_i\in [-2^K+1,2^K-1]$", which seems to suggest a method to implement range proof. As it is quite vague and provides no further detail or citation and its implementation is missing in Protocol 2 in \cite{Wikstrom2005}, more work is needed to confirm and instantiate the author's suggestion.

As no novel range proof technique is proposed in \cite{Wikstrom2005}, it can only employ an existing range proof technique to implement proof of (\ref{e8.5}). We need to eliminate the possibility that the author has another range proof technique in his mind, but forgets to include it in the description of the shuffling protocol in \cite{Wikstrom2005} by carelessness. Let's examine whether any of the existing range proof techniques fits the shuffling scheme in \cite{Wikstrom2005}. 

The most straightforward range proof technique is ZK proof of partial knowledge \cite{Cramer1994}, which proves that the committed integer may be each integer in the range one by one and then link the multiple proofs with OR logic. 
It has a drawback: low efficiency.
It is well known that it can be optimised by sealing the committed integer bit by bit and proving each commitment contains a bit. However, the optimised solution is still inefficient especially when the range size is large.

Boudot notices that $g^x$ is a computationally binding commitment of $x$ in $Z$ when the order of $g$ is unknown. Thus a computationally binding commitment function can be designed and any non-negative committed integer can be proved to be non-negative by showing that it is no smaller than a square. So Boudot employs commitment of integers in $Z$ instead of the traditional commitment of integers with a modulus in his range proof \cite{Boudot2000}. This special commitment function enables him to reduce range proof in a range $R$ to range proofs easier to implement. As a result his scheme achieves asymptotical precision. 
Moreover, it is more efficient than the solution based on \cite{Cramer1994}. 

Commitment of integers in $Z$ is employed in \cite{Lipmaa2003A} as well, which combines it with an interesting fact: any non-negative integer can be written as the sum of four squares. The range proof scheme in \cite{Lipmaa2003A} employs an algorithm to find the four squares to sum up any non-negative integer and then prove that the integer is non-negative through a proof of knowledge of the square roots of the four squares. 
In this way, it implements range proof and achieves completely perfect precision. Moreover, it has a constant cost independent of the range size. 

As the range proof techniques in \cite{Boudot2000} and \cite{Lipmaa2003A} are still not efficient enough, a very efficient range proof is widely used.  It is not systematically proposed in any single paper, but employed in a wide range of applications \cite{Bao1998,Fujisaki1998,Camenisch1998,Boudot1999,Ateniese2000,Poupard2000,Damgard2002,Camenisch2003,Camenisch2004,Kiayias2004,Choi2005,Kiayias2005,Kiayias2006,Wikstrom2005,Ge2006A,Ge2006B,Zhu2007}. The idea is simple: to prove that a secret integer $x$ is in a range $R$, a monotone function of $x$, $cx+r$ in $Z$ is published. If $cx+r$ is in a range $R'$, $x$ can be guaranteed to be in $R$. 
We call this method monotone test. Unlike the other range proof techniques, monotone test is not a general solution to range proof and its application has some special limitations (e.g. in choice of $R$, $R'$, $x$ and other parameters) as pointed out in \cite{Boudot2000} (Details like discussion of expansion rate can be found in \cite{Boudot2000}).

Which range proof technique may the author plan to employ to implement his shuffling scheme in \cite{Wikstrom2005}? We can deduce that it must be monotone test due to the following reasons.
\begin{itemize}
	\item It slightly modifies an already employed proof of knowledge of discrete of logarithm without employing any additional operation, so fits the author's suggestion ``proof of knowledge over a group of unknown order also gives an upper bound on the bit-size of the exponents, i.e., it implicitly proves ......"; while the other range proof technqiues needs explicit and addtional operations and should have been explictly described in Protocol 2 in \cite{Wikstrom2005} if employed.
	\item Only monotone test meets the efficiency claim of \cite{Wikstrom2005}  (in Section 5.5 of \cite{Wikstrom2005}) as it needs no extra cost. The other range proof technqiues are too costly for \cite{Wikstrom2005}. Even \cite{Boudot2000}, the most efficient of them, needs $40n$ more exponentiations to implement the $N$ range proofs in (\ref{e8.5}) and will contradict the efficiency claim of \cite{Wikstrom2005} and make it the least effcient shuffling scheme in the past decade.
\end{itemize}


If the $N$ instances of missing range proof in the shuffling scheme in \cite{Wikstrom2005} is implemented with monotone test, calculation of $d_i$ in Step 6 of Protocol 2 in \cite{Wikstrom2005} must be modified into 
$$
d_i=cp_{\pi(i)}+r_i \mbox{ in } Z
$$
such that $d_i$ becomes a modulus-free monotone function of $\rho_i=p_{\pi(i)}$. 
Namely, as mentioned before the original shuffling scheme in \cite{Wikstrom2005} must be modified into MP2 to fit monotone test.
Then in Step 7 of MP2, $d_i$ must be verified in a range $R'$. 
However, there are still two unclear but very important details to consider.
\begin{itemize}
	\item In \cite{Wikstrom2005}, as a very important security parameter, $K$ is not defined and its relation to other parameters is completely unknown. Obviously, (\ref{e2}) is not always guaranteed by (\ref{e8.5}), (\ref{e9}) and (\ref{e10}) with any $K$. To guarantee satisfaction of (\ref{e2}), $K$ must be appropriately instantiated. 
We believe that $K$ should be set as $K_3$, such that (\ref{e8.5}), (\ref{e9}) and (\ref{e10}) can guarantee (\ref{e2}) due to the following reasons.
\begin{itemize}
	\item If $K>K_3$, when there exist $i$, $j$ and $k$ such that $\rho_i=p_jp_k$, (\ref{e8.5}) can still be satisfied. For example, suppose $p_j$ and $p_k$ are the smallest primes larger than $2^{K_3-1}$, then it is very possible that their product is smaller than $2^K$. So when $K>K_3$, (\ref{e2}) cannot be guaranteed and thus soundness of the shuffling scheme in \cite{Wikstrom2005} fails.
	\item If $K<K_3$, when a $p_{\pi(i)}$ is larger than $2^K-1$ but smaller than $2^{K_3}$, the $\rho_i$ generated according to the shuffling protocol is larger than $2^K-1$ and thus cannot pass (\ref{e8.5}), which means an honest shuffling node cannot pass the verification and correctness of the shuffling scheme in \cite{Wikstrom2005} fails.
\end{itemize}
However, our deduction is not hinted or supported in any way in \cite{Wikstrom2005}.
	\item An appropriate $R'$ must be chosen to guarantee satisfaction of (\ref{e8.5}). However, there is not any hint in \cite{Wikstrom2005} about its existence or choosing method. As monotone test is not a general solution to range proof and is limited in application, it is unsure whether an appropriate $R'$ can be found. 
\end{itemize}

To confirm our deduction and clarify the mysteries above, we contacted the author of \cite{Wikstrom2005}.
In correspondence with us, the author confirmed our deduction and suggested to add a verification to Step 7 of Protocol 2 in his scheme to implement monotone test: $d_i<2^{K_3+K_4+K_5}$. He argues that this additional verification can guarantee that no $\rho_i$ can be large enough to be the product of two $p_i$s. Although the original Protocol 2 is not consistent with monotone test, which requires to calculate $d_i$ in a modulus-free way, MP2 proposed in Section~\ref{fc} can employ monotone test as it has modified the calculation $d_i$ to be modulus-free.
We combine MP2 and the additional test $d_i<2^{K_3+K_4+K_5}$ suggested by the author and denote the result as MSBMT (modified shuffling based on monotone test), where (\ref{e8.5}) is instantiated into
\begin{equation}\label{e8}
-2^{K_3}+1\le \rho_i\le 2^{K_3}-1 \mbox{ for } i=1,2,\dots,N.
\end{equation}

\subsection{Failure of the Author's Suggestion to Achieve Soundness}

Soundness of MSBMT lies in that the author believes that $\rho_i$ is guaranteed to be in $[-2^{K_3}+1, 2^{K_3}-1]$ when $d_i<2^{K_3+K_4+K_5}$.
Is his claim reliable? 
As monotone test is not a general solution we need to check whether it is suitable for the range proof in (\ref{e8}).
It is determined by whether the shuffling node can win the following game.
\begin{enumerate}
	\item When running MSBMT the shuffling node chooses $r_i$ and $\rho_i$ such that $\rho_i\notin [-2^{K_3}+1, 2^{K_3}-1]$.
	\item In the course of MSBMT a value is randomly chosen in $[2^{K_4-1},2^{K_4}-1]$ for the challenge $c$.
	\item The shuffling node outputs $d_i=c\rho_i+r_i$ to satisfy (\ref{e'0}), (\ref{e'0.5}), (\ref{e'1}) and (\ref{e'2}).
	\item He wins the game if $d_i$ happens to be smaller than $2^{K_3+K_4+K_5}$.
\end{enumerate}
If the shuffling node can win the game with a non-negligible probability, he can pass the verification of MSBMT with invalid $\rho_i$ with a non-negligible probability and thus soundness of MSBMT fails. Unfortunately, the shuffling node can always win the game using the following attack.
\begin{enumerate}
	\item When running MSBMT the shuffling node chooses an integer $\rho_i$ larger than $2^{K_3}-1$ and then a random integer $r_i$ smaller than $2^{K_3+K_4+K_5}-\rho_i2^{K_4}$. 
	\item When the shuffling node receives an integer $c$ in $[2^{K_4-1},2^{K_4}-1]$, he outputs $d_i=c\rho_i+r_i$.
\end{enumerate}
As $r_i<2^{K_3+K_4+K_5}-\rho_i2^{K_4}$, it is always satisfied that 
$$
d_i=c\rho_i+r_i<2^{K_4}\rho_i+r_i<2^{K_4}\rho_i+2^{K_3+K_4+K_5}-\rho_i2^{K_4}=2^{K_3+K_4+K_5}
$$
and he can always win the game.
Namely, no matter whether $\rho_i$ is in $[-2^{K_3}+1, 2^{K_3}-1]$ or how large its absolute value is, no matter which value is chosen from $[2^{K_4-1},2^{K_4}-1]$ as $c$ after the shuffling node chooses $\rho_i$ and $r_i$, he can always choose a special $r_i$ to pass the verification in MSBMT.
So (\ref{e8}) is not guaranteed in MSBMT.
When a $\rho_i$ is no smaller than $2^{K_3}$ it may be the product of multiple $p_i$s. 
When a $\rho_i$ is no smaller than $2^{K_3+1}-1$ it can certainly be the product of multiple $p_i$s without any doubt.  Therefore, soundness of MSBMT is compromised by the attack.

Maybe the author is careless and forgets to set the lower bound for $R'$. 
Maybe when he says $d_i$ must be smaller than $2^{K_3+K_4+K_5}$, he actually means $d_i$ must be in $[0,2^{K_3+K_4+K_5}-1]$. In this case, the attack above cannot work. However, another attack against soundness can work as follows.
\begin{enumerate}
	\item The shuffling node chooses an invalid positive value for $\rho_i$.
	\item He chooses $r_i\in [-2^{K_4-1}\rho_i,2^{K_3+K_4+K_5}-1-(2^{K_4}-1)\rho_i]$.
	\item When being challenged with $c$, the shuffling node returns a response $d_i=c\rho_i+r_i$.
\end{enumerate}
As 
\begin{eqnarray}
& d_i=c\rho_i+r_i=c\rho_i+r_i\le (2^{K_4}-1)\rho_i+2^{K_3+K_4+K_5}-1-(2^{K_4}-1)\rho_i= \nonumber \\
& 2^{K_3+K_4+K_5}-1, \nonumber \\
& d_i=c\rho_i+r_i=c\rho_i+r_i\ge 2^{K_4-1}\rho_i-2^{K_4-1}\rho_i=0, \nonumber
\end{eqnarray}
the attack can succeed on the condition that $-2^{K_4-1}\rho_i\le 2^{K_3+K_4+K_5}-1-(2^{K_4}-1)\rho_i$ and $[-2^{K_4-1}\rho_i,2^{K_3+K_4+K_5}-1-(2^{K_4}-1)\rho_i]$ is a valid range. As any $\rho_i$ with an absolute value no smaller than $2^{K_3}$ is invalid and $2^{K_5}$ must be overwhelmingly large (e.g. $K_5=50$ in \cite{Wikstrom2005}), this condition can be satisfied even if $\rho_i$ is much larger than $2^{K_3}$.

\subsection{Formal Proof of Infeasibility to Fix the Problem}

Although soundness of MSBMT fails with the $R'$ suggested by the author of \cite{Wikstrom2005}), optimistic opinion may still hope that the failure is caused by inappropriate choice of $R'$ and there may exist an appropriate $R'$ for monotone test to guarantee soundness of MSBMT.
However, Theorem~\ref{rg} illustrates that as application of monotone test is not general but limited no such $R'$ exists and it is impossible to limit the range of $\rho_i$ by checking the size of $d_i$ in MSBMT. 

\begin{theorem}\label{rg}
Suppose the absolute value of $\rho_i$ can be guaranteed to be smaller than $2^{K_3}$ in MSBMT when $d_i=cp_{\pi(i)}+r_i$ is in a certain range $[A,B]$, then a contradiction can be found.
\end{theorem}
\textit{Proof:}
\begin{itemize}
	\item On one hand, the width of $[A,B]$ must be smaller than $2^{K_3}(2^{K_4}-1-2^{K_4-1})$, otherwise $B-A\ge 2^{K_3}(2^{K_4}-1-2^{K_4-1})$ and the shuffling node can attack by choosing an invalid $\rho_i$ as $\rho_i=2^{K_3}$ and then $r_i\in [A-2^{K_4-1}2^{K_3},B-(2^{K_4}-1)2^{K_3}]$. In this attack, with any challenge $c$ in $[2^{K_4-1},2^{K_4}-1]$,  
\begin{eqnarray}
& d_i=c\rho_i+r_i=c2^{K_3}+r_i\le (2^{K_4}-1)2^{K_3}+B-(2^{K_4}-1)2^{K_3}=B, \nonumber \\
& d_i=c\rho_i+r_i=c2^{K_3}+r_i\ge 2^{K_4-1}2^{K_3}+A-2^{K_4-1}2^{K_3}=A. \nonumber
\end{eqnarray}
So the attack succeeds if $A-2^{K_4-1}2^{K_3}\le B-(2^{K_4}-1)2^{K_3}$ and $[A-2^{K_4-1}2^{K_3},B-(2^{K_4}-1)2^{K_3}]$ is a valid range.
As $B-A\ge 2^{K_3}(2^{K_4}-1-2^{K_4-1})$, 
\begin{eqnarray}
& (B-(2^{K_4}-1)2^{K_3})-(A-2^{K_4-1}2^{K_3})=(B-A)-(2^{K_4}-1-2^{K_4-1})2^{K_3} \nonumber \\
& \ge 2^{K_3}(2^{K_4}-1-2^{K_4-1})-(2^{K_4}-1-2^{K_4-1})2^{K_3}=0. \nonumber 
\end{eqnarray}
So $A-2^{K_4-1}2^{K_3}\le B-(2^{K_4}-1)2^{K_3}$ and $[A-2^{K_4-1}2^{K_3},B-(2^{K_4}-1)2^{K_3}]$ is a valid range, and thus the attack can succeed.
	\item On the other hand, when the width of $[A,B]$ is smaller than $2^{K_3}(2^{K_4}-1-2^{K_4-1})$ the probability that an honest shuffling node can pass the monotone test is negligible as shown in the following.
\begin{enumerate}
	\item The honest shuffling node has $\rho_i=p_{\pi(i)}$.
	\item The honest shuffling node chooses $r_i$ randomly from $[0,2^{K_3+K_4+K_5}-1]$.
	\item When being challenged with $c$, the honest shuffling node returns a response $d_i=c\rho_i+r_i$.
\end{enumerate}
As $r_i$ is randomly chosen from $[0,2^{K_3+K_4+K_5}-1]$, it is uniformly distributed in $[0,2^{K_3+K_4+K_5}-1]$. So no matter how large $c$ is, $d_i$ is uniformly distributed in a range as wide as $2^{K_3+K_4+K_5}$.
As the width of $[A,B]$ is smaller than $2^{K_3}(2^{K_4}-1-2^{K_4-1})$ and $2^{K_5}$ must be overwhelmingly large (e.g. $K_5=50$ in \cite{Wikstrom2005}), $d_i$ is uniformly distributed in a very large range, in comparison with which the width of $[A,B]$ is negligible. 
So the probability that $d_i$ falls in $[A,B]$ is negligible.
\end{itemize}
\hfill $\Box$ \\

\subsection{Unsuitability of Other Range Proof Techniques}
\label{ot}

Monotone test has been formally illustrated to be unable to implement the range proof in the original shuffling scheme in \cite{Wikstrom2005} or MSBMT. So the only way to obtain soundness is to apply another range proof technique to MP2 although they are not the choice of the author in \cite{Wikstrom2005}. Unfortunately, the other existing range proof primitives are much less efficient than monotone test. The $N$ instances of range proof in $[-2^{K_3}+1,2^{K_3}-1]$ (or at least in $[2^{K_3-1},2^{K_3}-1]$) cost $O(N2^{K_3})$ exponentiations with the original technique in \cite{Cramer1994} or $O(NK_3)$ exponentiations with its optimisation and both are much more costly than the cost of the original shuffling scheme in \cite{Wikstrom2005}. Although the range proof primitives in \cite{Boudot2000} and \cite{Lipmaa2003A} are more efficient, application of $N$ instances of either of them is still too costly for shuffling. As both of them can only work with special commitment functions, to employ either of them firstly the shuffling node must commit to each secret logarithm to be proved in the certain range in a special commitment function and then he has to prove that the same logarithm is used in the shuffling protocol and committed in the commitment function using zero knowledge proof of equality of discrete logarithms (\cite{Chaum1992} or its variant). After that the range proof primitive in \cite{Boudot2000} or \cite{Lipmaa2003A} can be applied and each application cost several exponentiations. Not only the cost of the $N$ instances of range proof but also the commitment functions and their corresponding proof of equality of logarithms are more costly than the original shuffling scheme in \cite{Wikstrom2005}. So employing any other range proof than monotone test will greatly increase the cost in computation and communication and leads to an shuffling scheme much less efficient than most existing shuffling schemes.

\section{Conclusion}

The mix network in \cite{Wikstrom2005} fails in correctness, has an unreliable proof of zero knowledge and cannot achieve soundness. Although we managed to restore correctness by fixing the proof protocol and adopting statistical zero knowledge, soundness is impossible in \cite{Wikstrom2005} as a key operation in it, $N$ instances of range proof, cannot be implemented.
As a result, a simple attack can compromise the shuffling scheme.
In the shuffling protocol in \cite{Wikstrom2005}, not only range proof is not implemented, but also the suggested method cannot implement it.
Fixing the suggested range proof is impossible, while no alternative technique can guarantee soundness of the shuffling scheme in \cite{Wikstrom2005} at a tolerable cost. That may be why implementation of range proof is only very vaguely hinted and completely missing in the detailed implementation in \cite{Wikstrom2005} supposed to be complete although it is very important. 
Our conclusion is that if we cling to monotone test suggested by the author there is no way to guarantee soundness. If it is replaced by any other range proof techniques \cite{Cramer1994,Boudot2000,Lipmaa2003A}, the additional operations in the $N$ instances of range proof will greatly increase the cost in computation and communication and leads to an shuffling scheme much less efficient than most existing shuffling schemes.

\appendix

\section{Drawbacks of [39] and [28]}
\label{dr}

There are two drawbacks in \cite{Peng2004A}. Firstly, \cite{Peng2004A} only supports a small fraction of all the possible permutations. So its privacy is much weaker than that of other shuffling schemes, which support all the possible permutations. Secondly, it requires that the shuffling node (e.g. tallier) does not obtain collusion of the message providers (e.g. voters) and has no knowledge of the shuffled messages. So it is impractical in applications with critical requirement on security like political e-voting, whose soundness cannot rely on any trust on the participants.

The shuffling scheme in \cite{Groth2007} employs exceptionally small parameters to improves efficiency of \cite{Furukawa2005}, so its soundness is weaker.
Firstly, it employs much smaller challenges in its ZK proof than in \cite{Furukawa2005}, so its soundness may fail with a much larger (although still regarded small in circumstances with looser security requirements) probability.
Secondly, it depends on bindingness of a commitment function but achieves too weak bindingness in the commitment function.
Commitment function $com(m_1,m_2,\dots,m_k,r)=\prod_{i=1}^kg_i^{m_i}h^r$ is employed in \cite{Furukawa2005} where the order of $g_1$, $g_2$, $\dots,g_k$ and $h$ is $q$.    
Soundness of the three-move proof in \cite{Groth2007} is based on an assumption: binding property of the commitment function is computationally unbreakable such that in the third move the prover is forced to use the unique set of secret integers committed in the first move to generate his response to the random challenge in the second move (see the two theorems in \cite{Groth2007} for more details of this assumption). 
It is illustrated and emphasized in \cite{Groth2003,Wikstrom2005} (which employ the same main idea) and recognised in \cite{Groth2007} that in such proof mechanism soundness of shuffling fails if the commitment is not binding and the prover can adjust the committed integers after it receives the challenge.
In \cite{Groth2007} the length of $q$ is only 240 bits for the sake of high efficiency. Note that obtaining $\log_{g_i}g_j$ or $\log_{g_i}h$ is enough to break the binding property of the commitment function, which depends on hardness of calculating such discrete logarithms. Therefore, breaking soundness of the shuffling in \cite{Groth2007} is no harder than calculating a 240-bit discrete logarithm. 
As a building block claimed to be mainly used in electronic voting (which is often a very sensitive political activity and requires very high level of security), it is inappropriate for the shuffling protocol to base its security on hardness to calculate a 240-bit discrete logarithm, which is commonly regarded to be not hard enough for a powerful adversary in the current security standard. In comparison, the other shuffling schemes employ bases with much larger orders and base their soundness on hardness to calculate at least 1024-bit long discrete logarithms.
In summary, \cite{Groth2007} drastically improves efficiency of \cite{Furukawa2005} by using exceptionally small parameters. However, it weakens  soundness to an extent intolerable in many e-voting applications.

\end{document}